# Pulsar spins from an instability in the accretion shock of supernovae

John M. Blondin[1] & Anthony Mezzacappa[2]

[1]Department of Physics, North Carolina State University, Raleigh, North Carolina 27695-8202, USA.
[2]Physics Division, Oak Ridge National Laboratory, Oak Ridge, Tennessee 37831-6354, USA.

**Rotation-powered radio pulsars are born with inferred initial rotation periods[1] of order 300 ms (some as short as 20 ms) in core-collapse supernovae. In the traditional picture, this fast rotation is the result of conservation of angular momentum during the collapse of a rotating stellar core. This leads to the inevitable conclusion that pulsar spin is directly correlated with the rotation of the progenitor star[2]. So far, however, stellar theory has not been able to explain the distribution of pulsar spins, suggesting that the birth rotation is either too slow[3] or too fast[2,4]. Here we report a robust instability of the stalled accretion shock in core-collapse supernovae that is able to generate a strong rotational flow in the vicinity of the accreting proto-neutron star. Sufficient angular momentum is deposited on the proto-neutron star to generate a final spin period consistent with observations, even beginning with spherically symmetrical initial conditions. This provides a new mechanism for the generation of neutron star spin and weakens, if not breaks, the assumed correlation between the rotational periods of supernova progenitor cores and pulsar spin.**

The collapse of a massive star's core that triggers a supernova explosion is followed by a brief epoch of less than a second during which the nascent supernova shock wave stalls at a radius of order 100 km and is revived, and the supernova initiated, by an as yet undetermined mechanism[5]. Hydrodynamics simulations have shown that this quasi-steady shock is subject to the stationary accretion shock instability, or SASI[6–9]. However, these two-dimensional simulations admit only axisymmetric modes and, hence, the resulting dynamics cannot affect the rotation of the accretion flow. As we show here, in three dimensions non-axisymmetric modes can significantly alter the angular momentum of the collapsed core.

We have performed on a three-dimensional cartesian grid a series of simulations of a steady accretion shock, following the numerical approach described in ref. 6 and in the Supplementary Information. We found that the nonlinear evolution of the SASI is dominated by a low-order non-axisymmetric mode characterized by a spiral flow pattern beneath the accretion shock. The SASI has been interpreted in terms of a growing acoustic wave propagating around the periphery of the shocked accretion flow—that is, around the periphery of the region between the proto-neutron star (PNS) and accretion shock, or post-shock region[10]. Whereas the axisymmetric sloshing mode (characterized by $l = 1$ in spherical harmonics) seen in earlier work represents the propagation of this wave along a symmetry axis from one pole to the other, the spiral mode ($m = 1$) discovered here represents the propagation of this wave around an axis, as illustrated in Fig. 1.

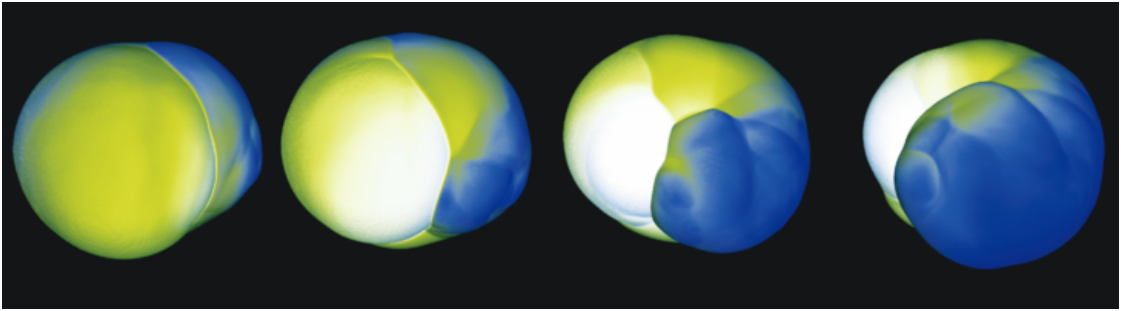

**Figure 1** | **The evolution of the supernova accretion shock illustrates the rotation of the spiral mode of the SASI.** The blue portion of the shock surface represents the leading portion of the spiral SASI wave, seen here propagating from right to left across the front face of the shock. The discontinuity between the blue and white surfaces is the shock triple point marking the leading edge of the SASI wave. An animation of this evolution is available in Supplementary Information.

In the nonlinear regime this SASI wave propagating around the inside surface of the accretion shock creates two strong counter-rotating flows as seen in Fig. 2. Over the leading half of the SASI wave, the gas immediately behind the accretion shock is moving in the same direction as the propagation of the wave. This flow is fed in part by the obliquity of the accretion shock, which refracts the radially falling gas above the shock into a rotational motion moving with the SASI wave. As the leading edge of the SASI wave travels around the accretion shock, it subverts the weaker, receding portion of the accretion shock ahead of it and drives the lower-entropy shocked gas interior to the receding portion of the shock down towards the PNS. The orientation of the weak accretion shock in this region leads to a post-shock flow moving in a direction opposite to that of the SASI wave.

The spiral mode of the SASI is a robust result of a stalled accretion shock in three dimensions. We have evolved a dozen simulations with different initial perturbations, from random acoustic noise in the accretion shock cavity to various configurations of density perturbations in the infalling material above the shock. In all cases the late-time evolution was dominated by this spiral mode. We also ran three simulations (using different initial perturbations) with moderate rotation of the infalling gas to model the effect of a rotating progenitor star, using a specific angular momentum at the accretion shock comparable to the 15-solar-mass model described in ref. 4. In the presence of progenitor rotation, the spiral SASI becomes dominant much more quickly than in the absence of rotation.

The spiral flow pattern generated by the distorted accretion shock will have a marked effect on the underlying PNS. Figure 3 shows the time evolution of the net angular momentum accreted onto the PNS. For the non-rotating-progenitor models there is no angular momentum in the flow entering the simulation domain at the outer boundary, nor are there any external torques that might change the global angular momentum. Therefore the net angular momentum of the simulation must remain zero. As a consequence, the angular momentum in the accretion flow above the surface of the PNS should be equal and opposite to the angular momentum of the accreting PNS. Our simulation maintains this equality to within a few per cent, with the difference attributable to numerical errors inherent in advecting angular momentum on a cartesian grid. The separation of angular momentum plotted in Fig. 3 is a direct result of the spiral SASI wave, which generates two counter-rotating flows as described above. The net result is that gas with angular momentum of one sign is accreted onto the PNS, whereas gas with the opposite angular momentum flows around the outer regions of the shock cavity, presumably driven outwards once the supernova explosion is initiated. The most striking feature in Fig. 3 is the almost linear increase in the PNS angular momentum about 1 s after the simulation begins, and even earlier for a rotating progenitor.

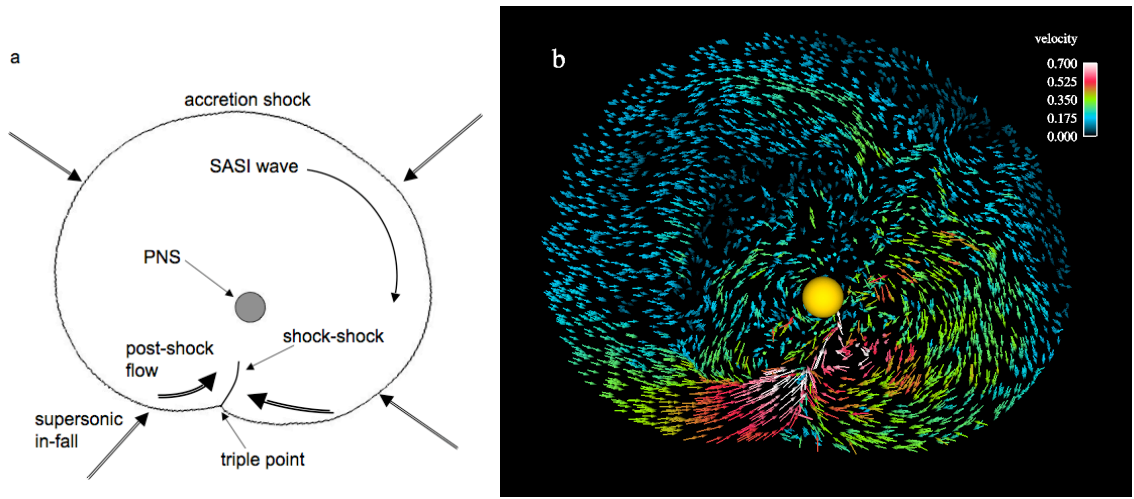

**Figure 2 | The flow in the equatorial plane of the spiral SASI mode drives accretion of angular momentum onto the PNS. a**, This diagram illustrates the shock structure and corresponding post-shock accretion flow created by the spiral SASI wave. The location of the accretion shock is taken from the equatorial plane of a three-dimensional simulation with the shock pattern (the SASI wave) propagating in a clockwise direction. The leading edge of the internal SASI wave is marked by a shock-shock[13]: a shock wave formed by the steepening of a pressure wave propagating along the inside surface of the accretion shock. This shock-shock connects to the accretion shock at a triple point, seen as a discontinuity in the surface of the accretion shock. In three dimensions this triple point is a line segment on the surface of the accretion shock that spans roughly half the circumference, as seen in Fig. 1. **b**, The flow vectors highlight two strong rotational flows. On the right the flow is moving clockwise along with the shock pattern, whereas at the bottom left the post-shock flow is being diverted into a narrow stream moving anticlockwise, fuelling the accretion of angular momentum onto the PNS.

The presence of rotation in the infalling gas helps to initiate the spiral mode of the SASI, as demonstrated by the early increase in angular momentum seen in Fig. 3 and the fact that the rotation axis of the SASI wave is roughly aligned (by 10°, 15° and 45°) with the spin axis of the progenitor star. This alignment has the surprising effect of erasing the progenitor spin from the PNS. Because the angular momentum accretion driven by the SASI is opposite to that of the SASI wave itself, the initial effect is to spin down the PNS. The comparable rate of increase in accreted angular momentum between the non-rotating and rotating models shows that the magnitude of the angular momentum accretion rate is set by the flow pattern of the spiral SASI wave, not by the angular momentum of the infalling gas above the accretion shock.

**Figure 3 | The spin-up of the accreting PNS due to the spiral SASI mode.**

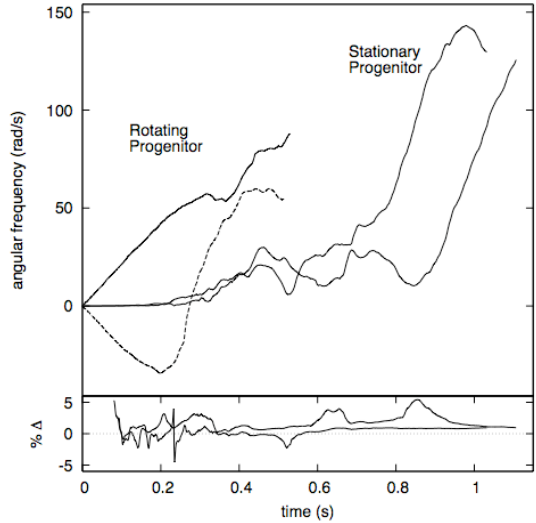

The total accreted angular momentum (divided by the moment of inertia of an isolated neutron star such that the unit of measure is the spin frequency of the relic neutron star) is plotted as a function of time for three three-dimensional SASI simulations with different initial perturbations. The bottom panel shows the difference between the angular momentum of the PNS and the total angular momentum in the flow above the surface of the PNS for the non-rotating-progenitor models. Their vector sum should, in fact, cancel given that we began with spherically symmetrical initial conditions and, hence, no net angular momentum. For the rotating-progenitor model we also show the $x$ component of the angular momentum (dashed line). At early times the accreted angular momentum is dominated by the progenitor rotation (aligned along the $-x$ axis), but once the SASI becomes nonlinear it dominates the angular momentum accretion rate, and the rotation of the accreted material reverses direction.

The net angular momentum accreted onto the PNS as a result of the spiral SASI mode during the supernova initiation phase can markedly alter the spin rate of the neutron star left behind. Over a timespan of 250 ms the PNS in our models will typically accrete 0.1 solar masses and a net angular momentum of $2.5 \times 10^{47}$ g cm$^2$ s$^{-1}$. If we divide this total angular momentum by the moment of inertia of an isolated neutron star[11], $I \approx 2 \times 10^{45}$, we find that the resulting rotational period would be about 50 ms. (We use the results from ref. 12 to scale our models to relevant supernova values: a shock radius of 230 km, 1.2 solar masses interior to the shock, and a mass accretion rate of 0.36 solar masses per second.) This does not include any relic angular momentum from the progenitor core rotation. Thus, for an initially non-rotating or slowly (angular momentum about $2.5 \times 10^{47}$ g cm$^2$ s$^{-1}$ or less) rotating progenitor the SASI will be the dominant source of angular momentum in the remnant neutron star and will produce a spin rate consistent with the inferred birth periods of pulsars[1]. By comparison, current stellar evolution models[4] predict a relic angular momentum of at least $8 \times 10^{47}$ g cm$^2$ s$^{-1}$, which would result in a birth period of about 15 ms. Such fast rotation is marginally consistent with only the fastest radio pulsars. Either current predictions of progenitor core rotation are too large or some unknown mechanism would serve to spin down the PNS (the SASI would reduce the relic angular momentum by only about 30%). If the former is true and core rotation rates are actually significantly lower, the final spin period of the neutron star would be determined by the SASI, as discussed above.

The precise final spin period of the remnant neutron star will be determined by the length of time for which the SASI spiral mode is dominant, which will depend both on how fast it is initiated after core bounce and how long the stalled accretion shock persists before an explosion is initiated. The former is affected by the (poorly known) progenitor rotation, whereas the latter will require three-dimensional supernova models with sufficient realism to follow the entire explosion process—for example with three-dimensional, multi-frequency neutrino transport (the models considered here are valid only for the stalled-shock phase). The different simulations shown in Fig. 3 illustrate the uncertainty in our models of the time at which the spiral flow pattern begins but at the same time confirm the robustness of the spin-up induced by the spiral SASI. The outcomes provide a new mechanism for the generation of neutron star spin. Moreover, they demonstrate that progenitor spin and neutron star spin will not be as simply correlated as previously believed.

**Supplementary Information** is linked to the online version of the paper at www.nature.com/nature.

**Acknowledgements** This work was supported by a SciDAC grant from the US Department of Energy High Energy, Nuclear Physics, and Advanced Scientific Computing Research Programs. A.M. is supported at the Oak Ridge National Laboratory, managed by UT-Battelle, LLC, for the US Department of Energy. The simulations presented here were performed at the Leadership Computing Facility at ORNL. We thank the National Center for Computational Sciences at ORNL for their resources and support.

**Author Information** Reprints and permissions information is available at www.nature.com/reprints. The authors declare no competing financial interests. Correspondence and requests for materials should be addressed to J.M.B. (john_blondin@ncsu.edu).